\documentclass[aps,prc,preprint,groupedaddress,showpacs,preprintnumbers,
amsmath,amssymb]{revtex4}

\usepackage{graphicx}
\usepackage{dcolumn}
\usepackage{bm}

\begin{document}


\title{
On the Accuracy of Hyperspherical Harmonics Approaches
to Photonuclear Reactions
}


\author{Nir Barnea$^1$,  Winfried Leidemann$^2$, 
Giuseppina Orlandini$^2$, Victor D. Efros$^3$ Edward L. Tomusiak$^4$}
\affiliation{$^1$ The Racah Institute of Physics, The Hebrew University, \\ 
91904 Jerusalem, Israel \\
$^2$Dipartimento di Fisica, Universit\`a di Trento\\
and Istituto Nazionale di Fisica Nucleare, Gruppo Collegato di Trento, \\
I-38050 Povo, (Trento) Italy \\
$^3$Russian Research Centre ``Kurchatov Institute'', \\
Kurchatov Square, 1, 123182 Moscow, Russia \\
$^{4}$Department of Physics and Astronomy, University of Victoria, \\
Victoria, BC V8P 1A1, Canada}


\date{\today}

\begin{abstract}
Using the Lorentz Integral Transform (LIT) method we compare the results for the triton
total photodisintegration cross section obtained 
using the Correlated Hyperspherical Harmonics (CHH) and the Effective Interaction Hyperspherical Harmonics (EIHH) techniques. We show that these two approaches, while rather different both conceptually and computationally, lead to results which 
coincide within high accuracy. The calculations which 
include two-- and three--body forces are of the same high quality in both cases.
We also discuss the comparison of the two approaches in terms of computational efficiency.
These results are  of major importance in view of applications to the much debated
case of the four--nucleon photoabsorption. 
\end{abstract}

\pacs{21.45.+v, 25.10.+s, 25.20.DC, 27.10.+h}


\maketitle


\section{\label{Introduction} Introduction}
In the last few years there has be a resurgence of interest
in the photodisintegration of few--nucleon systems. 
This is essentially due  to two reasons.
One of them is that few--body theories have now achieved the ability
to produce results of high accuracy for reactions into the many--body continuum.
The other is the ongoing debate about the existence and nature of three--body forces,
which finds a natural testing ground in reactions involving the three--nucleon systems. 
Some theoretical results on photonuclear reactions with light nuclei using realistic 
two-- and three--body potentials have already 
been published \cite{ELOT:2000,GSGKNWELOT:2002,GSWGNK:2005} 
and among them is a benchmark between the correlated hyperspherical harmonics (CHH) 
and Faddeev approaches \cite{GSGKNWELOT:2002} on the total photodisintegration cross
section of the three--body systems. This benchmark demonstrated the high degree of
accuracy reached by these 
two techniques. At the same time it showed the inadequacy of  
existing experimental data to allow any conclusion about the three--nucleon force effect, whose magnitude 
turns out to be smaller than the rather large  error bars.


In this work we test the EIHH method  on the triton photonuclear cross section with realistic forces.
This method is also based on a hyperspherical harmonics expansion, but uses the concept of an effective interaction to speed up the convergence of the expansion. 
We believe that it is necessary to benchmark EIHH results on a realistic force calculation.
In fact, when combined with the LIT method \cite{ELO:1994}, this approach is very promising
(see e.g. \cite{BMBLO:2002,BABLO:2004}) for the study of electromagnetic reactions on heavier systems, where three--body force effects  might be more prominent.

In the following we first briefly recall the LIT method (Section \ref{LIT}). In Section \ref{HH} the hyperspherical expansion method is summarized, together with the CHH (Section \ref{CHH}) and EIHH (Section \ref{EIHH}) techniques. They address the problem of convergence for the solutions of the
Schr\"odinger and  Schr\"odinger--like  equations required by the LIT method. 
Section \ref{Results} contains results for the triton photoabsorption with a realistic two-- and three--nucleon force and, in view of applications to the electromagnetic responses of heavier systems,
also a discussion about the computational efficiency of the two approaches.

\section{\label{LIT} The LIT method}

The LIT method has been described extensively in several publications \cite{ELO:1994, ELO:1999}. 
Here we only recall the key points.

The calculation of any inclusive electromagnetic cross section requires the knowledge 
of the response function
\begin{equation}\label{response}
R(\omega)=\sum\!\!\!\!~\!\!\!\!\!\!\!\!\int _n\,\,|\langle 0|\Theta|n\rangle|^2\delta(\omega-E_n+E_0) \, ,
\end{equation}
where $\omega$ represents the energy transferred by the elctromagnetic 
probe,  $|0\rangle$ and
$E_0$ are ground state wave function and energy of the system undergoing 
the reaction, $|n\rangle$ and $E_n$ denote eigenstates and eigenvalues of 
the Hamiltonian $H$, and $\Theta$ is the operator relevant to the
reaction. 
The relation between the total photoabsorption cross section and the response function is
given by
\begin{equation}
\sigma_\gamma(\omega)= 4 \pi^2 \alpha \omega R(\omega)\,,
\end{equation}
where $\alpha$ is the fine structure constant and 
$\Theta$ of Eq. (\ref{response}) is the nuclear current operator. 
In the low--energy region considered here one
can rely on Siegert's theorem and use for $\Theta$ the unretarded dipole operator
\begin{equation}\label{operator}
\Theta=\sum_i^A z_i \tau_i^3\,.
\end{equation}

The LIT method consists in calculating $R(\omega)$ in three steps.
{\it Step 1.} 
The equation  
\begin{equation} \label{LITeq}
(H- E_0-\omega_0 + i \Gamma)| \tilde \Psi \rangle = \Theta |0\rangle\,.
\end{equation}
is solved for many $\omega_0$ and a fixed $\Gamma$.

This is a Schr\"odinger--like equation with a source. It can be shown easily 
that the solutions $|\tilde\Psi \rangle$ are localized.
Thus  one only needs a bound state technique to obtain them. In particular one
can adopt the same bound state technique as for the solution of the ground state, which
is an input for Eq.~(\ref{LITeq}).

The values of the parameters $\omega_0$ and $\Gamma$ are chosen in relation 
to the physical problem. In fact, as will become clear in {\it Step 2}, the 
value of $\Gamma$ is a kind of ``energy resolution'' for the response function
and the values of $\omega_0$ scan the region
of interest. In our case of triton photodisintegration
we solve Eq.~(\ref{LITeq}) with $\Gamma=10$ and 20 MeV 
and for a few hundred of $\omega_0$ values chosen in the interval 
$-40$ MeV to $200$ MeV. 

{\it Step 2.} The overlaps 
$\langle\tilde\Psi|\tilde\Psi\rangle$ of the solutions from {\it Step 1.} are
calculated. Of course these overlaps
depend on $\omega_0$ and $\Gamma$. A theorem for integral transforms based on the closure property 
of the Hamiltonian eigenstates \cite{EFROS:1985} ensures that in the present case this dependence 
can be expressed as \cite{ELO:1994}
\begin{equation}\label{deflit}
 {\rm L} (\omega_0,\Gamma)=\langle\tilde\Psi|\tilde\Psi\rangle= \int R(\omega)\, 
          {\mathcal L}(\omega,\omega_0,\Gamma)\, 
d\omega\,,
\end{equation} 
where ${\mathcal L}$ is the Lorentzian function centered at
$\omega_0$ and with $\Gamma$ as a width:
\begin{equation}
{\mathcal L}(\omega,\omega_0,\Gamma)=
  {1\over (\omega -\omega_0)^2 + \Gamma^2}\,.
\end{equation}
Therefore by solving Eq.~(\ref{LITeq}) one can easily obtain the Lorentz integral
transform of the response function. 

{\it Step 3.} The transform (\ref{deflit})
is inverted (see e.g. Ref. \cite{ELO:1999,ALRS:2005})
in order to obtain the response function  and with it the
cross section. Of course the inversion result has to be independent of $\Gamma$ and
show a high degree of stability.

\section{\label{HH} The HH expansion technique}

In the HH expansion approach the basis functions for the solution of the 
$A$--body Schr\"odinger equation
for the ground state $|0\rangle$, as well as of the $A$--body 
``LIT equation'' (\ref{LITeq}) for $|\tilde\Psi\rangle$,
are expressed as functions of the hyperspherical coordinates. 
These are constructed starting from 
the $A-1$ Jacobi coordinates $\vec\xi_i$ for the $A$--particle system. 
They consist of a hyperradius $\rho_{A-1}=\sqrt{\sum_i^{A-1}(\vec\xi_{i})^2}$ 
and a hyperangle 
$\Omega_{A-1}=\{\hat{\xi}_1,\hat{\xi}_2,\cdots,\hat{\xi}_{A-1},
              \varphi_2,\varphi_3,\cdots,\varphi_{A-1}\}$ 
$=\{\Omega_{A-2},\varphi_{A-1},\hat{\xi}_{A-1}\}$, where the 
$A-2$ hyperangles $\{\varphi_2,\varphi_3,\cdots,\varphi_{A-1}\}$ are 
defined in the $[0,\pi/2]$ interval through the relation
\begin{equation}  \sin{\varphi_n}=\xi_{n}/\rho_n\,,\quad n=2,3,\cdots,A-1\,.
\end{equation}
In terms of these coordinates the
Laplace operator for $n$ Jacobi coordinates is 
\begin{equation}
  \Delta_n=\frac{1}{\rho_n^{3n-1}}\frac{\partial}{\partial\rho_n}
           \rho_n^{3n-1}\frac{\partial}{\partial\rho_n}
          -\frac{1}{\rho_n^2}\hat{K}_n^2\,,
\end{equation}
i.e. a sum of a hyperradial operator and of a term containing the hyperspherical, 
or grand angular momentum operator $\hat{K}_n^2$. The latter can be expressed 
recursively in terms of the hyperangular momentum operator associated with 
the subset $\{\vec{\xi}_{1},\vec{\xi}_{2},\cdots,\vec{\xi}_{n}\}$, 
$\hat K^2_{n-1}$, and $\hat\ell_{n}^2$, the orbital momentum operator 
corresponding to the $n$th Jacobi coordinate, namely~\cite{EFROS:1972}: 
\begin{equation}
   \hat{K}_n^2=-\frac{\partial^2}{\partial\varphi_n^2}
               +\frac{3n-6-(3n-2)\cos{2\varphi_n}}{\sin{2\varphi_n}}
\frac{\partial}{\partial\varphi_n}
\frac{1}{\cos^2{\varphi_n}}\hat{K}_{n-1}^2
+\frac{1}{\sin^2{\varphi_n}}\hat{\ell}_{n}^2.
\end{equation}
In the above equation $\hat{K}_1^2\equiv\hat{\ell}^{\,2}_{1}$, and the angular
momentum operator associated with these $n$ coordinates is 
$\vec{\hat{L}}_n=\vec{\hat{L}}_{n-1}+\vec{\hat{\ell}}_{n}$. 
The operators $\hat{K}_n^2$, $\hat{\ell}_{n}^{\,2}$, $\hat{K}_{n-1}^2$, 
$\hat{L}_n^2$ and $\hat{L}^{(z)}_n$ commute with each other.  

Given the set of  hyperspherical  coordinates, the hyperspherical harmonics 
functions ${\mathcal Y}_{[K_n]}$ are the eigenfunctions of the hyperangular
momentum operator $\hat{K}^2_n$,
\begin{equation}
\hat{K}^2_n{\mathcal Y}_{[K_n]}=K_n(K_n+3n-2){\mathcal Y}_{[K_n]}\,,
\end{equation}  
where $K_n$ is the hyperangular quantum number and the symbol $[K_n]$ 
stands for the set of quantum numbers 
$\{K_n,K_{n-1},\cdots,K_2;L_n,M_n;L_{n-1},\cdots,L_{2}
;\ell_{n},\ell_{n-1},\cdots,\ell_{1}\}$. 
The explicit expression for the HH functions of $n$ Jacobi coordinates 
can be constructed recursivly as described in
Ref.~\cite{EFROS:1972}.

The HH functions ${\mathcal Y}_{[K_n]}$ form a complete and orthonormal 
set of functions that satisfy
\begin{equation}
\left\langle{\mathcal Y}_{[K_n]}\left.\right|
            {\mathcal Y}_{[K^{\,\prime}_n]}\right\rangle
           =\delta_{[K_n],[K^{\,\prime}_n]}\,.
\end{equation} 
The particle permutation properties of the HH functions can be implemented in 
different ways.
In the CHH approach used here this is done as described in Ref. \cite{FE:1981}, 
while in the EIHH framework
two recently developed powerful algorithms are employed \cite{BN:1997,BN:1998}.


As functions of the hyperspherical coordinates the antisymmetric
spin--isospin configuration space
basis functions are written in the form
\begin{equation}\label{HHexpansion}
   {\mathcal{HH}}_{N[K,\gamma]}^{[f]}(\rho,\Omega_{A-1},\sigma,\tau)
    =R_N(\rho)\Gamma_{[K,\gamma]}^{[f]}(\Omega,\sigma,\tau)\,,
\end{equation}
where, for the hyperradial part of the wave function one chooses an 
orthonormalized set of functions, 
$R_N(\rho)\sim L_N^\nu(\rho/\rho_0)\exp(-\rho/2\rho_0)$.
Here $L_N^\nu$ are the generalized Laguerre polynomials, and $\nu$ and $\rho_0$ 
are free parameters ~\cite{FE:1981}. 
In the fully antisymmetric functions $\Gamma_{[K,\gamma]}^{[f]}$ the notation $[K,\gamma]$ stands for the set of quantum numbers $\{K,J,M_J,L,S,T,M_T\}$, i.e. total angular momentum and its  third component,  orbital angular momentum,  spin, total isospin and isospin projection of the A--body system, respectively and $\sigma,\tau$ indicate collectively the spin and isospin variables of the $A$ nucleons. These functions are obtained by the coupling of the HH functions ${\mathcal Y}_{[K]}^{[f]\mu_f}(\Omega)$ with specified $L$ and $M$ of the representation $[f]$ and the spin--isospin functions $\Theta_{SM_STM_T}^{[\bar{f}]\mu_{f}}(\sigma,\tau)$ with specified $S, M_S, T$ and $M_T$ of the conjugate Young scheme $[\bar{f}]$.
In many cases the rate of convergence of the HH expansion is rather slow.
In order to accelerate it one can follow
two rather different approaches, leading to the CHH and EIHH methods.

\subsection{\label{CHH} The CHH Approach }
The CHH approach \cite{FE:1972} has been used extensively in the past. 
While the underlying idea is common, there exist a number of slightly
different versions of this approach. In the following we describe how
the CHH has been applied by the present authors in this and previous works
  \cite{ELOT:2000,ELOT:2004}. 
  
The main idea of the CHH approach consists of the insertion in Eq.~(\ref{HHexpansion})
of a Jastrow factor 
${\mathcal J}$ embodying the short range correlation due to the repulsive part of the potential. 
Such a repulsion leads to high momentum components in the wave function which make
the convergence of the  HH expansion rather slow. Therefore one modifies 
Eq.~(\ref{HHexpansion}) into 
\begin{equation}\label{CHHexpansion}
{\mathcal{CHH}}_{N[K,\gamma]}^{[f]}(\rho,\Omega_{A-1},\sigma,\tau)={\mathcal J}\,R_N(\rho)\Gamma_{[K,\gamma]}^{[f]}(\Omega,\sigma,\tau)\,.
\end{equation}
Here ${\mathcal J}$ is taken as a state--dependent correlation operator of the form
\begin{equation}
{\mathcal J}={\mathcal S}\prod_{i<j}\sum_{s,t}{f_{s,t}(r_{ij})}\hat P_{s,t}(ij)\,,
\end{equation}
where $\hat P_{s,t}(ij)$ are projection operators onto nucleon pairs $(ij)$ with spin $s$ and isospin $t$
and ${\mathcal S}$ is a particle symmetrization operator. The radial parts of the correlation operator are chosen as follows. For $r_{ij}<r_0$ (the healing distance) $f_{s,t}(r_{ij})$ is chosen to be the zero energy pair wave function in the corresponding ${s,t}$ state. The healing is insured by imposing the condition  
$f_{s,t}(r_{ij})=1$ for $r_{ij} > r_0$ and $f'_{s,t}(r_{ij})=0$ for $r_{ij}=r_0$. The ${s,t}={1,3}$ and 
${s,t}={3,1}$ cases are determined from the $^1S_0$ and $^3S_1$ partial wave of the NN potential.
Since for the ${s,t}={1,1}$ and 
${s,t}={3,3}$ cases the $^1P_1$ and $^3P_1$ potentials are not sufficiently attractive to obtain a healing distance, we introduce an additional intermediate range central interaction. Further details can be found in \cite{LEOT:1999}.
\subsection{\label{EIHH} The EIHH approach} 
In contrast to the CHH approach, which focuses on the short range properties of the wave functions,
the EIHH approach focuses on the potential operator. The acceleration of the convergence
is obtained by introducing  a two--body 
effective interaction $V^{[2]}_{eff}$ \cite{BLO:2000,BLO:2001}. This is done by the division of 
the total HH space in P and Q spaces realized via the HH quantum number $K$ (P space: 
$K \leq K_{max}$, Q space: $K > K_{max}$ ). Since 
in hyperspherical coordinates the total Hamiltonian is written as
\begin{equation}
  H =\frac{1}{2 m} \left(-\Delta_{\rho}+ \frac{\hat{K}^2}{\rho^2}\right) 
         + \sum_{i<j} V_{ij} \;,
\label{H}
\end{equation}
one can extract from it a ``two--body'' Hamiltonian
of particles $A$ and $(A-1$)
\begin{equation}
  H_{2}(\rho) = \frac{1}{2 m} \frac{\hat{K}^2}{\rho^2} 
         + V_{A,(A-1)} \;,
\label{HH2}
\end{equation}
which, however, contains the hyperspherical part of the A--body kinetic energy. 
Since the HH functions of the $(A-2)$ system are eigenfunctions of 
$\hat K^2_{A-2}$ one has an explicit dependence of $H_2$ on the quantum 
number $K_{A-2}$ of the residual system, i.e. $H_{2} \rightarrow 
H_{2}^{K_{A-2}}$.
Applying the hermitian version of the Lee--Suzuki method \cite{LS:1980,SUZUKI:1982,SO:1983}
to $H_2$ one gets an effective Hamiltonian $H_{2eff}$. The effective 
interaction $V^{[2]}_{eff}$ is obtained from 
\begin{equation}
 V_{eff}^{[2] K_{A-2}}(\rho) = H_{2eff}^{K_{A-2}}(\rho) - \frac{1}{2 m} 
\frac{\hat{K}^2}{\rho^2}\,.
\label{HH3}
\end{equation}
This $V^{[2]}_{eff}$ replaces $V_{ij}$ in Eq.~(\ref{H}) when one projects the solution
on the P--space.
This effective potential has the following property: $V^{[2]}_{eff}\to V_{ij}$ for 
$P\to 1$. Due to the ``effectiveness'' of the operator the solution of the 
Schr\"odinger equation converges faster to the true one. 
The present HH formulation of an effective potential has the following peculiarities, which make 
it particularly ``effective'': (i) $V^{[2]}_{eff}$  is $\rho$ dependent, therefore it contains 
 information on the ``medium''; (ii) because of the above mentioned $K_{A-2}$ dependence
the (A-2) residual system is not a pure spectator and its state influences 
the state of the pair, therefore $V^{[2]}_{eff}$ can be viewed as a state dependent effective 
interaction. Moreover we point out that the effective  potential 
can be interpreted as a kind of momentum expansion, since the 
short range resolution is increased with growing $K_{max}$. 
\section{\label{Results} Results}
\begin{figure}
\resizebox*{15cm}{17cm}{\includegraphics[angle=-90]{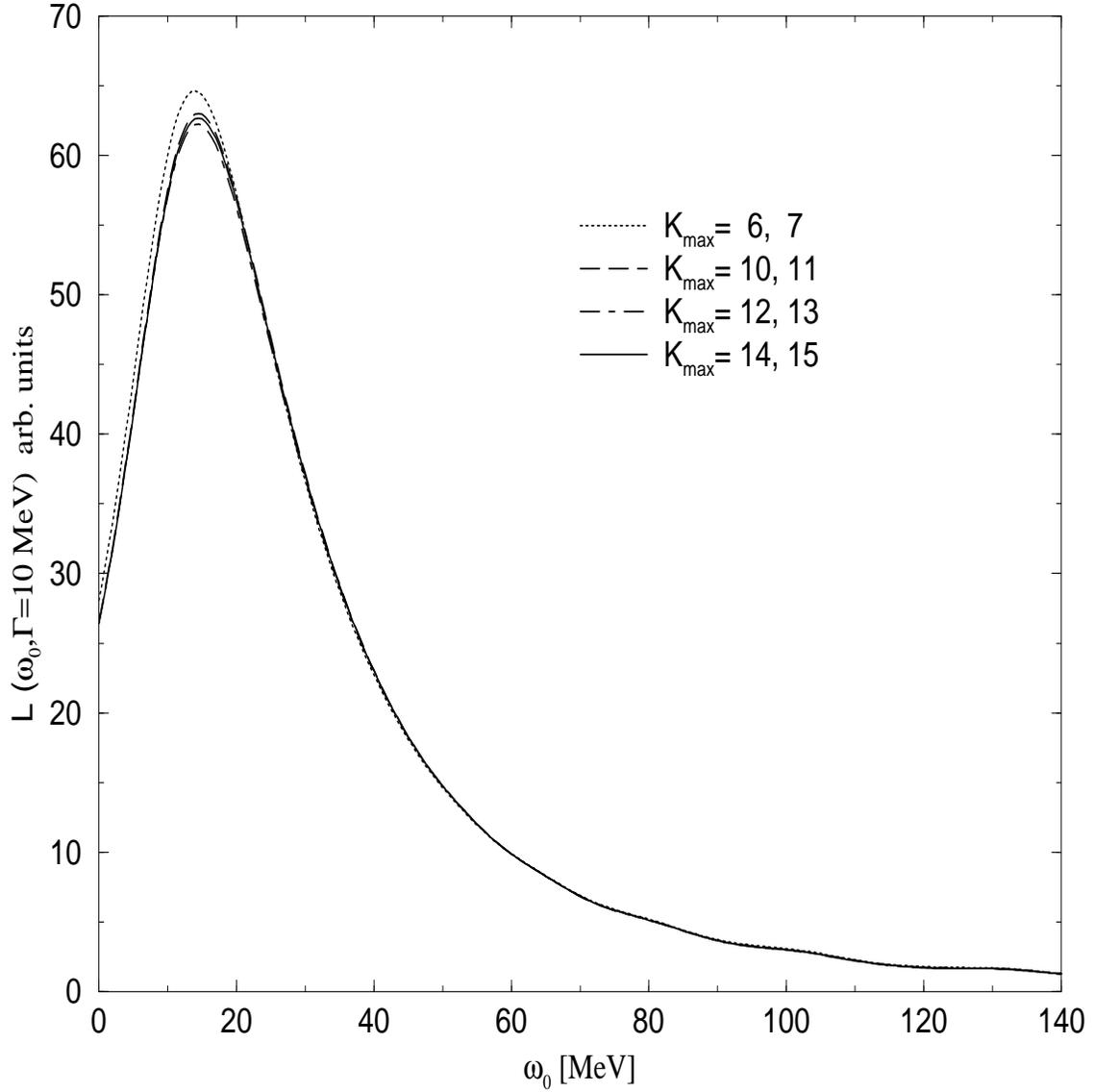}}
\caption{Rate of EIHH convergence of the LIT of the unretarded dipole response function
(Eq.(\ref{deflit})) for $\Gamma=10$ MeV, limited to the (J=1/2,T=1/2) component. 
The even values of $K_{max}$ refer to the calculation of the ground state $|0\rangle$, 
the odd values to that of the Lorentz state $\tilde\Psi$.} 
\label{figure1}
\end{figure}

\begin{figure}
\resizebox*{15cm}{17cm}{\includegraphics[angle=0]{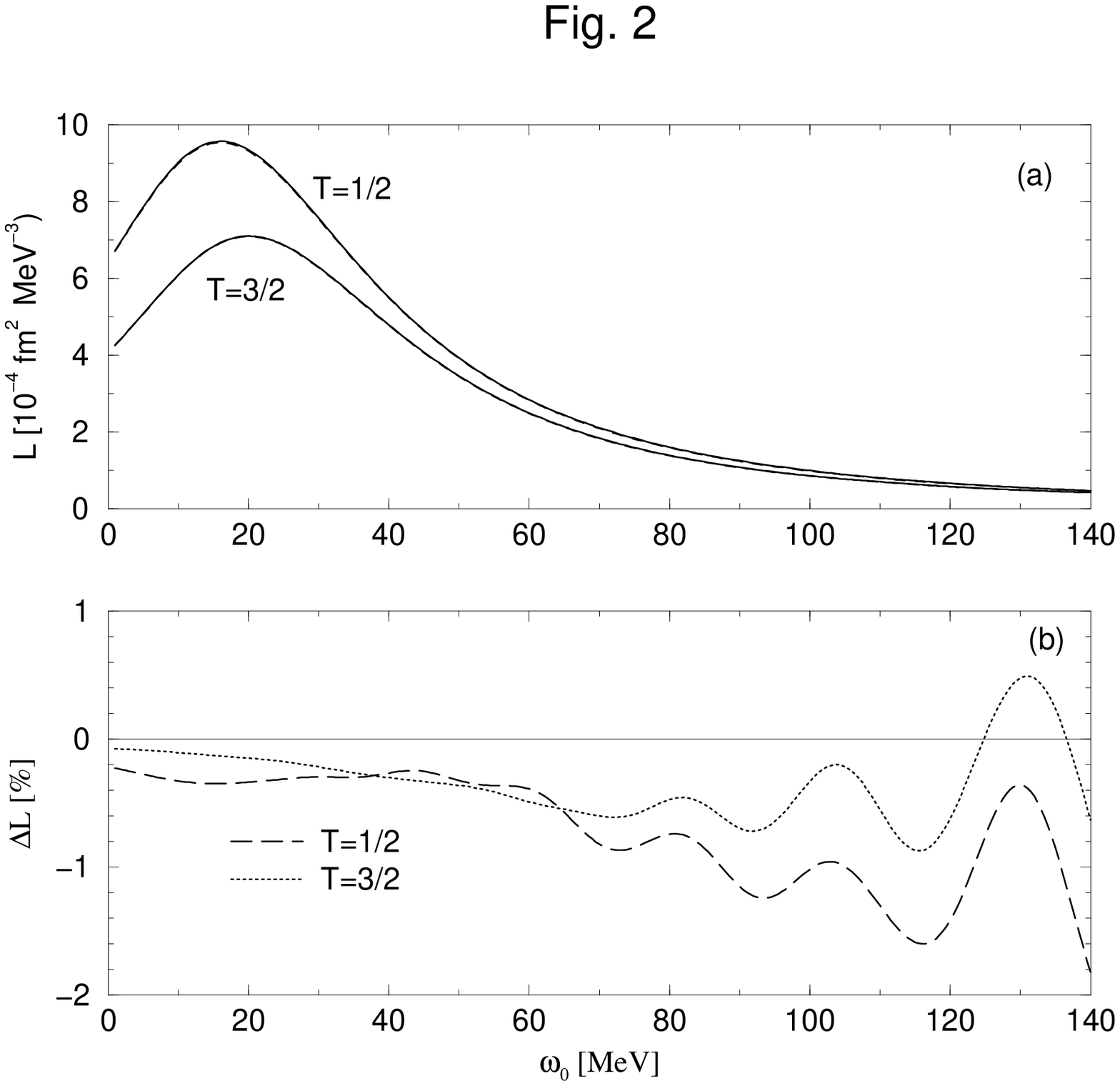}}
\caption{Comparison between CHH and EIHH results for the LIT
of the unretarded dipole response function (Eq.(\ref{deflit})) 
in the two isospin channels. (a):  CHH (full curve); EIHH (dashed curve).
(b): The difference $\Delta {\rm L}= ({\rm L}^{\rm EIHH}- {\rm L}^{\rm CHH})
/{\rm L}^{\rm CHH}$ between the CHH and EIHH results of Fig.~2a expressed in percent.}
\label{figure2}
\end{figure}

\begin{figure}
\resizebox*{15cm}{17cm}{\includegraphics[angle=0]{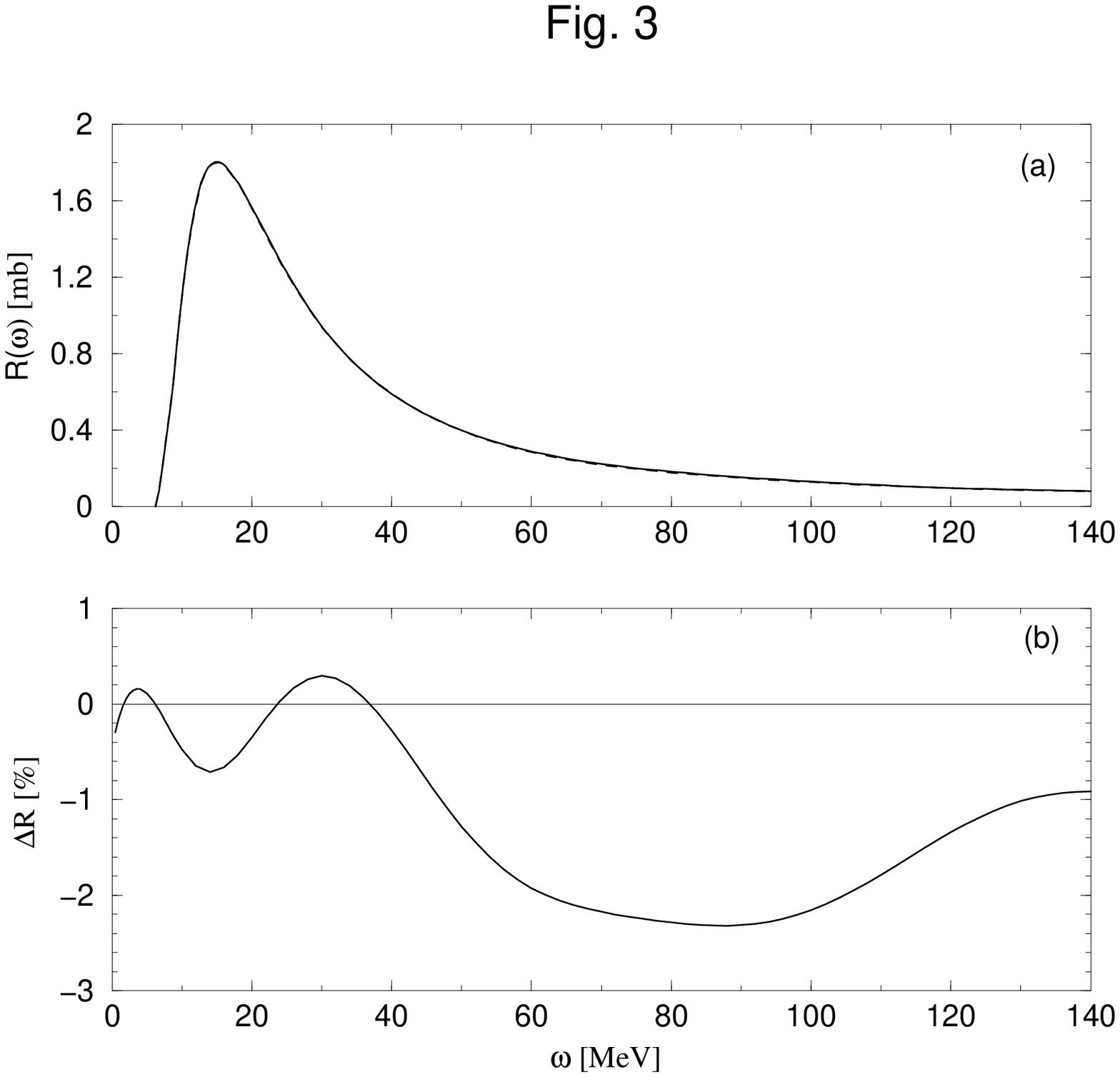}}
\caption{Comparison between CHH and EIHH total photonuclear cross sections of triton.
(a): CHH (full curve); EIHH (dashed curve), (b): The difference $\Delta R= ( R^{\rm EIHH}- R^{\rm CHH})/R^{\rm CHH}$ between the CHH and EIHH results of Fig.~3a expressed in percent.} 
\label{figure3}
\end{figure}

We have compared the transform $ {\rm L}(\omega_0, \Gamma)$ and
response function $R(\omega)$ as computed by the CHH and 
EIHH approaches.  The potentials employed were the
AV18 \cite{AV18:1995} two--body and
UIX \cite{UBIX:1997} three--body potential models.
In the case of CHH we expand the ground state with 11 hyperradial functions and 
the hyperspherical states with $K$ up to $K_{max}=10$, but consider
additional selected states up to $K_{max}=100$. 
For the T=1/2 component of $|\tilde\Psi\rangle$ we take 26 hyperradial 
functions, $K_{max}=9$ and additional selected states up to $K_{max}=65 $. 
For T=3/2 $K_{max}=15$ and 26 hyperradial functions are sufficient for a convergent result. 
In the case of EIHH we expand with 26 hyperradial functions up to 
$K_{max}=14$  (ground state) and $K_{max}=15$ ($|\tilde\Psi\rangle$). 
In Fig.~\ref{figure1} the EIHH rate of convergence of $ {\rm L}(\omega_0, \Gamma)$ is 
shown for the (J=1/2,T=1/2) contribution to $\tilde\Psi$ with $\Gamma=10$ 
MeV. One sees that the convergence is very good in the whole energy range. 
One also notes that small oscillations are present in the transform at higher
energies. If needed one could improve the result by increasing the number of 
hyperradial states. Convergence patterns of the other (J,T) 
contributions in the EIHH calculation are similar to that shown
in Fig.~\ref{figure1}. Also the rate of convergence of the CHH calculation for T=3/2 is 
rather similar to that of Fig.~\ref{figure1}, while for T=1/2 the convergence is 
considerably slower. Comparable differences in the rate of convergence
of the two methods have also been found for the $^4$He total photoabsorption cross section 
with semirealistic potential models \cite{BELO:2001}. Thus, in view of applications 
to more complex nuclei, the EIHH method seems to be preferable from the calculational 
point of view due to its faster convergence. 
In Fig.~\ref{figure2}a  we show the results for the total Lorentz integral transform of $R(\omega)$ 
(contributions of J=1/2 and J=3/2 components are summed) for both the  T=1/2 and 
T=3/2 cases and with $\Gamma=20$ MeV. It is evident that
for both isospin channels the agreement is excellent over  the whole considered energy range.
In Fig.~\ref{figure2}b the relative difference 
$\Delta {\rm L}= ({\rm L}^{\rm EIHH}- {\rm L}^{\rm CHH})
/{\rm L}^{\rm CHH}$ is shown in percent. It is readily seen that $\Delta {\rm L}$ 
 does not exceed 0.4\% in the peak region. For higher energies the
differences can become somewhat larger but remain below 2\% up to pion threshold.
In Fig.~\ref{figure3}a we show the total photoabsorption cross section. The results of the two 
approaches remain very close also after inverting the transforms ({\it Step 3} in 
Section \ref{LIT}). In fact the EIHH and CHH curves are hardly distinguishable. Even 
if the agreement has slightly deteriorated compared to 
Fig.~\ref{figure2} due to the inversion procedure, the quality remains excellent.
As shown in Fig.~\ref{figure3}b the difference remains  below 0.8\% 
in the peak region and has a maximum of 2.4\% up to pion threshold. 
We would like to point out that the agreement of the two HH calculations in the
peak region is even better than the already good agreement of CHH and Faddeev 
calculations shown in the above
mentioned benchmark \cite{GSGKNWELOT:2002}. We do not compare  the present results with 
the Faddeev cross section in \cite{GSGKNWELOT:2002} since that benchmark was made neglecting the charge dependent terms in the AV18 nuclear force. Though that effect is small (of the order of 1\% in the peak region), it is much larger
than the difference between CHH and EIHH calculations.
If one considers that EIHH and CHH approaches differ substantially both from the conceptual and calculational
points of view (of course completely independent codes are used), the agreement of the results is
really remarkable and shows that one can rely on the 
application of the EIHH approach to heavier systems with a high degree of confidence. 
In particular, its application  
to the calculation of the total photonuclear cross section of $^4$He within realistic nuclar force
models, containing two-- and three--body forces, would be very much needed.
It could shed some light on the role of three--nucleon forces, which might
be even more important than already found in the three--nucleon system. 
Moreover, it would make a valuable contribution to the clarification 
of the present rather confused situation in the comparison between theoretical and experimental results
in $^4$He (see e.g. \cite{MAXLAB:2005,JAPAN:2005}). 
Work in this direction is in progress.
\section*{Acknowledgment}
This work was supported by the grant COFIN03 of the Italian Ministery of University 
and Research. V.D.E. acknowledges support from the RFBR, grant 05-02-17541. 
The work of N.B. was supported by the ISRAEL SCIENCE FOUNDATION while E.L.T.
was supported by the National Science and Engineering Research Council of
Canada.
(grant no. 202/02). 
\newpage
\bibliography{bib}

\newpage

\end{document}